\documentclass[final,5p,times,twocolumn,authoryear]{elsarticle}

\usepackage{amsmath, mathrsfs, amssymb,amsfonts,amsthm,graphicx, epsf, dcolumn, yfonts}
\usepackage{booktabs}
\usepackage{array} 
\usepackage{hyperref}
\usepackage{cuted}
\usepackage{color}
\usepackage{orcidlink}
\DeclareMathAlphabet{\mathpzc}{OT1}{pzc}{m}{it}
\definecolor{MyDarkRed}{rgb}{0.71,0.14,0.07}
\newcommand{\edd}[1]{{\color{black}{#1}}}

\def\Z#1{_{\lower2pt\hbox{$\scriptstyle#1$}}}
\def\goesas{\mathop{\sim}\limits}
\def\lsim{\mathop{\hbox{${\lower3.8pt\hbox{$<$}}\atop{\raise0.2pt\hbox{$\sim$}}$}}}

\newcommand{\ave}[1]{\left\langle #1 \right\rangle}

\newcommand{\order}[1]{ \mathcal{O} \left( #1 \right) }

\def\dd{\mathop{\text{d}\!}}

\def\Md{M_\text{d}} \def\rd{r_\text{d}} 
\def\Mb{M_\text{gc}} \def\Rb{r_\text{gc}}

\journal{Physics Letters B}

\begin{document}

\begin{frontmatter}

\title{The need for a nonlocal expansion in general relativity}

\author[first]{Marco Galoppo\orcidlink{0000-0003-2783-3603}}
\ead{marco.galoppo@canterbury.ac.nz}
\affiliation[first]{organization={University of Canterbury},
{School of Physical & Chemical Sciencies},
            addressline={Private Bag 4800}, 
            city={Christchurch},
            postcode={8140}, 
            state={},
            country={New Zealand}}
\author[second]{Giorgio Torrieri\orcidlink{0000-0002-0611-766X}}
\ead{torrieri@unicamp.br}
\affiliation[second]{organization={Universidade Estadual de Campinas},
{Instituto de Fisica ``Gleb Wataghin'',},
            addressline={Rua Sergio Buarque de Holanda 777 }, 
            city={Campinas},
            postcode={13083-859},
            state={},
            country={Brasil}}

\begin{abstract}
  Motivated by known facts about effective field theory and non-Abelian gauge theory, we argue that the post-Newtonian approximation might fail even in the limit of weak fields and small velocities for wide-extended rotating bodies, where angular momentum spans significant spacetime curvature. We construct a novel dimensionless quantity that samples this breakdown, and we evaluate it by means of existing analytical solutions of rotating extended bodies and observational data. We give estimates for galaxies and binary systems, as well as our home in the Cosmos, Laniakea. We thus propose that a novel effective field theory of general relativity might be needed to account for the onset of nonlocal angular momentum effects.
\end{abstract}


\end{frontmatter}

\section{Introduction}
The post-Newtonian approximation of general relativity (GR) is arguably the most successful effective field theory (EFT) in physics \citep{Will_2011, Porto_2016}. The post-Newtonian approximation has been applied both at astronomical, galactic and cosmological scales, with remarkable success throughout all of these scenarios \citep{Clifford_2014,Poisson_2014}. Nonetheless from the galactic scale onward the introduction of non-baryonic dark matter is required for the EFT predictions to match the observations.

The problem of dark matter \citep{Bertone_2005,Bekenstein_2004} is a major indicator that ``new physics''\ is required to understand the large scale structure of the universe.  Evidence ranging from disc galaxy rotation curves \citep{Rubin_1978,Bosma_1978,Sofue_2001}, to gravitational lensing events \citep{Treu_2010,Bartelmann_2010}, to the large structure of the universe \citep{DelPopolo_2007} suggests that ``gravity''\, and ``visible matter''\, do not match. This requires either a new source of matter or a modification to gravity in the weak-field limit.

Yet attempts have been made to suggest that general relativistic effects without Newtonian analogues, when properly accounted for, could explain the astrophysical observations at the galactic level. The proposed analytical solutions for stationary, axisymmetric extended disc galaxies in general relativity \citep{Cooperstock_2007,Balasin_2008,Astesiano_2022a,Astesiano_2022b,Ruggiero_2024,Re_2024,Galoppo_2024b} appear to be very far from the post-Newtonian limit.
Whilst the first models of this type were undoubtedly unphysical (e.g., by assuming rigid rotation \citep{Cooperstock_2007,Balasin_2008}), in recent years these models have shown vast improvement, by introducing differential rotation \citep{Astesiano_2022a,Astesiano_2022b,Ruggiero_2024,Re_2024} and even pressure \citep{Galoppo_2024b}. However, their net disconnection with any conventional post-Newtonian limit of disc galaxy dynamics has remained. This has made most of the research community sceptical about the physical relevance of such models \citep{Rowland_2015,Ciotti_2022,Lasenby_2023}: the gravitational field of galaxies is extremely weak w.r.t.\ the nonlinearities of GR, and the post-Newtonian expansion has been rigorously developed and tested (see e.g., \citep{Clifford_2014,Poisson_2014,Porto_2016}). While nonlinear general relativistic effects have been proposed to explain dark energy at cosmological scales \citep{Notari_2006,Wiltshire_cosmic_2007-1,Wiltshire_CEP_2008,Buchert_2012}, no convincing general reason has been put forward for why such effects should be present in the case of galaxies.

In this work, we argue however that the Post-Newtonian expansion could, in fact, fail even for weak gravitational fields and nonrelativistic velocities in a very particular case, i.e., that of extended bodies rotating over scales where the spacetime curvature changes. We note that this situation would then be relevant for galactic and super-galactic scales but not for systems of few bodies, even relativistic ones such as black hole pairs.

To build up the argument, we note that whilst effective field theory (EFT) development can be very mathematically sophisticated, its basic principles are relatively simple: one writes down all terms compatible with the fundamental symmetries of the system, and then orders them based on one, or more, small parameters, usually related to scale separation.  \textcolor{black}{In particular}, the post-Newtonian expansion \citep{Clifford_2014,Poisson_2014,Porto_2016} is based on two such scales, i.e., (i) the speed of light is ``large'', w.r.t.\ the typical speeds within the physical system; and (ii) the gravitational field strength is ``weak'', so that nonrelativistic particles cannot be accelerated to relativistic speeds. Then, these two scales are conventionally put together via a single expansion parameter, which relates the compactness of the system to the orbital velocities within it.  The fact that this applies to few-body scattering in classical field theory has been very well established both formally and in phenomenologically \citep{Porto_2016}.
\textcolor{black}{Indeed, \cite{Porto_2016} shows systematically that for few-body physics the expansion generated by the post-Newtonian EFT exactly parallels the Feynman diagram based expansion of quantum EFT used in particle physics scattering problems. Therefore, the post-Newtonian framework exactly provides a classical EFT whose perturbative structure mirrors that of Feynman-diagram expansions in quantum scattering problems.} 

\textcolor{black}{It is therefore worth examining in more detail what happens beyond the limit where this matching was well-studied.  In fact, while} the procedures to build EFTs are universal, there are some well-known cases in which the effective expansion fails \citep{Donoghue_2009}. For such an expansion to make sense, the ``fundamental'', theory must be more symmetric than the EFT for its low-energy limit. However, this is not always the case. When a conservation law present in the effective expansion is violated in the fundamental theory, then the effective expansion is dubious even when a well-defined scale of parameters exists. In quantum physics this becomes inevitable as the representations of the Hilbert space between the two theories are different, but even in classical physics, for periodic solutions, a small violation of a conservation law can build up over many cycles. For a small number of degrees of freedom the Kolmogorov–Arnold–Moser (KAM) theorem guarantees that perturbations to a system dynamics due to symmetry breaking terms remain negligible \citep{Goldstein_2002}. However, as the number of degrees of freedom goes to infinity this is no longer true \citep{Villani_2011}.  Thus, if both angular momentum and a large number of degrees of freedom are present, weak-field expansions are inherently suspicious.

Within field theory, a famous example of the reasoning above is the ``weak-field'', expansion of quantum chromodynamics (QCD). Superficially, it is useful for many applications to expand the chromo-fields around a classical background, $A_\mu \rightarrow A_\mu^{\mathrm{classic}} + \delta A_\mu$, as this approach can produce the correct Feynman rules and $\beta-$function \citep{Peskin_1995}.  However, an EFT that expands chromo-fields around a classical background in general will miss the fundamental feature of this theory, i.e., quark confinement.  The reasons for this are not fully understood, but fundamentally, we know that a Lorentz-covariant non-Abelian gauge can be set only locally in the space of field configurations \citep{Gribov_1978,Dudal_2008}.

Therefore, a ``Post-Maxwellian''\, expansion will be plagued by a multitude of configurations where potentials grow arbitrarily strong, but for which the action is still equivalent to a weak expansion of the fundamental theory. Similarly to the inverse cascade in turbulence \citep{Blaizot_2008}, the redundancies force the dynamics of colour chargers to be driven by larger and larger structures. This turbulent dynamics becomes quenched only when one assumes states separated by a certain distance in configuration space (related to the separation of the redundancies in field space) to be colour singlets, which can be thought of as imposing a higher symmetry on the infrared theory than on the fundamental one \citep{Pisarski_2006}. Thus, the correct EFT expansion in the regime where the theory is strongly coupled, but confinement has not been completely hidden in the fundamental building blocks of the theory, is not based on fields in QCD, but rather on nonlocal objects represented by the Wilson lines and loops \citep{Pisarski_2006}. 

Note that while this is usually thought of as a ``quantum effect" due to the strong coupling of the theory in the infrared. However, on a deeper investigation, such a stance would be tautological, since strong coupling is an artefact of the running of the coupling constant with distance. The real issue is the non-existence of a  ``semiclassical limit" compatible with the fundamental symmetries of the theory, where small perturbations, classical or quantum, are divided from the background.

Of course, the above should be meant only as a rough analogy. There is no confinement in GR as the theory is not defined by a tangent bundle coupling to a vector conserved current, and graviton-graviton interactions have a very different scale dependence than gluon-gluon interactions (the latter being marginal while the former are irrelevant). Observables in GR are not gauge invariant, and redundancies in asymptotic geometry play a very different role from a spin 1 gauge theory. Nevertheless, some subtleties of non-Abelian gauge theories are also present in GR, making the analogy not totally useless. In GR the ``gauge'' can only be set locally up to small deformations, but it keeps depending on the asymptotic structure of the spacetime. While GR is profoundly symmetric, its symmetries are local redundancies rather than global continuous symmetries that generate conservation laws via Noether's theorem \citep{Avery_2015}.  As such, angular momentum is exactly conserved in the classical Newtonian theory and special relativity, but not in GR. Within GR, angular momentum is crucially sensitive to the asymptotics of frame-dragging effects, and, in a general spacetime, it's not globally conserved due to curvature gradients \citep{Penrose_1982}. General covariance ensures local conservation of angular momentum in inertial frames, but by its nature it is not a local quantity, being defined over arbitrarily large distances w.r.t.\ a reference point (this is analogous to the proposed relation between turbulence and redundancy dominated dynamics in \citep{Blaizot_2008}).

The reasoning above makes it clear that it is possible that the post-Newtonian expansion could fail, even when its separation scales naively hold, for a many-body system that (i) is extended enough that the curvature varies significantly within its structure; (ii) carries a significant amount of angular momentum. Thus, in a many-body system, we would identify variations of relativistic angular momentum and curvature over comparable scales as cause for a possible break-down of the Newtonian expansion. We notice that this seems exactly to be the case when full GR solutions of extended rotating bodies are considered \citep{Astesiano_2022a,Re_2024,Galoppo_2024b}.

\section{The expansion parameter}
Given the discussion above, one can try to define a nonlocal Lorentz-scalar, to function as a dimensionless expansion parameter (according to Buckingham's $\pi$ theorem \citep{buckingham}) which quantifies the criteria above, namely a special relativistic angular momentum defined over regions of curvature variation. It is then convenient to work within a foliation framework, with a 3 + 1 splitting of a general spacetime \citep{Gourgoulhon_2012}, and define a Lorentz scalar ultimately adapted to the spatial hypersurfaces of the foliation. From dimensional analysis, the lowest-order, non-zero object of this type\footnote{To illustrate this, we could consider a linear parameter in $J^{\alpha \beta \gamma}$. The only non-zero quantity analogous to \eqref{eq:dalpha} would be $\order{m^{1}} \times J^{\alpha \beta \gamma} R_{\alpha \beta \gamma \nu} d\Sigma^\nu$. However, $J^{\alpha \beta \gamma}  d\Sigma^\nu$ is not easily relatable to a conserved quantity and there is no natural constant in GR $\sim m$.  These points can only be overcome by a quantity $\order{J \times J}$.} is (given a coordinate chart over the spacetime)
\begin{equation}   \delta \tilde{\alpha} (x^\mu,y^\mu) := G\,  R_{\alpha '\beta '\gamma\delta}(x^\mu,y^\mu) J^{\alpha ' \beta ' \xi'}(x^\mu) J^{\gamma \delta \zeta}(y^\mu) \dd \Sigma_{\xi '}(x^\mu) \dd \Sigma_\zeta (y^\mu)  \, , \label{eq:dalpha}
\end{equation}
where indices with a prime relate to the (co-)tangent space  at $x^\mu$, and regular indices to the one in $y^\mu$. Here, $G$ is the gravitational constant, $J^{\alpha \beta \gamma}$ is the orbital angular momentum tensor, $J^{\mu\nu\sigma}:= X^{[\mu}T^{\nu]\sigma}$, where \edd{$T^{\alpha\beta}$ is the effective energy-momentum tensor of the system under consideration,} $X^\mu$ is the four-position field of the matter elements \edd{(i.e., either individual particles positions in a discrete system or cell centres in a coarse-grained fluid description )}, and $\dd\Sigma^\mu$ is the timelike four-vector related to an infinitesimal three-volume element in the spacetime. \textcolor{black}{Note that, to obtain a dimensionless parameter from the physical quantities involved, we set $c = \hbar = 1$ (equivalent to defining the meter in terms of the second, and the second in terms of the Joule) and introduce a factor of $G$ (which sets the absolute scale, in inverse Planck scale units). Indeed, with this choice of units the $J^{\bullet\bullet\bullet}\times\dd \Sigma_\bullet$ terms are dimensionless, whilst $R_{\bullet\bullet\bullet\bullet}$ takes the units of a mass squared, i.e., the inverse units of $G$. Moreover, the power of $G$ matches that appearing in the Einstein--Hilbert action, making this a natural candidate for a power-counting expansion. Here, we note that in field theory, $G$ is interpreted as a dimensionful ultraviolet cutoff for gravity and is multiplied by the characteristic infrared scale to estimate the effective-theory expansion parameter. Therefore, in an astronomical context, this motivates computing the relevant quantities as $G M_{\odot}^{2}$ multiplied by a dimensionless integral.}

Moreover, we have defined the interaction kernel as a bi-tensor \citep{Synge_1960,Poisson_2011} via
\begin{align}
    R_{\alpha' \beta' \gamma \delta}(x^\mu,y^\mu):=& \frac{1}{2}\left[\mathcal{P}_{\alpha '\xi}(x^\mu,y^\mu)\mathcal{P}_{\beta '\zeta}(x^\mu,y^\mu)R^{\xi \zeta}_{\phantom{\xi} \phantom{\zeta} \gamma \delta}\left(y^\mu\right) \nonumber \right.
    \\
    & \left.+\mathcal{P}_{\gamma \xi'}(x^\mu,y^\mu)\mathcal{P}_{\delta \zeta '}(x^\mu,y^\mu)R_{\alpha ' \beta '}^{\phantom{\alpha '} \phantom{\beta '} \xi' \zeta'}\left(x^\mu\right)\right]\, .
\end{align}
Here, $\mathcal{P}_{\alpha'\beta}(x^\mu,y^\mu)$ is the canonical parallel transport bi-tensor from $(t;\vec{x})$ to $(t;\vec{y})$ along the shortest geodesic within the spatial hypersurface \citep{Poisson_2011}\footnote{We note that not for all spacetimes such geodesic is unique. When necessary, a direct generalisation of $\mathcal{P}_{\alpha'\beta}(x^\mu,y^\mu)$ is obtained via direct averages of the parallel transport along all such minimal geodesics. This average would then produce a well-defined expansion parameter, as its bi-tensors are also correctly defined.}, and $R_{\alpha\beta\gamma\delta}$ indicates the standard Riemann tensor. Then, by direct double integration of $\dd \tilde{\alpha}$ on the physical system volume, $V$, within a spacelike foliation slice we obtain
\begin{equation}
\label{eq:myalpha}
    \tilde{\alpha} = G\iint_{V\times V} R_{\mu'\nu'\rho\sigma}(x^\mu,y^\mu)J^{\mu'\nu'\beta'}(x^\mu)J^{\sigma\rho\gamma}(y^\mu)\dd\Sigma_{\beta'}(x^\mu)\dd\Sigma_{\gamma}(y^\mu)\, .
\end{equation}
This novel expansion parameter, although covariantly defined, depends on the choice of foliation\footnote{We also remark that in principle the expansion parameter in the post-Newtonian EFT is not only foliation-dependent but also, non-covariantly defined. However, it can be made covariant by employing the spatial projection on the chosen foliation to define the spatial velocity of the matter.}. Such a situation is not quite unique to this new quantity, as almost every scalar out of such a spatial integration would be foliation dependent \citep{Mourier_2024}. 

Moreover, we emphasise that the expansion parameter is defined on a system-by-system basis. For example, the value of $\tilde{\alpha}$ computed for a super-cluster characterises the dynamics of that system as a whole, and should not be expected to describe its smoothed-out components, such as individual galaxies or dust clouds. More generally, any estimation of $\tilde{\alpha}$ must be interpreted within the context of the specific physical system under consideration, as determined by the chosen coarse-graining scale and the corresponding effective matter description. This is particularly relevant in GR, where averaging procedures mix gravitational and matter degrees of freedom and can change the effective description of many-body systems \citep{Ellis_1984,Ellis_1987,Buchert_2000,Buchert_2001,Buchert_2006,Buchert_2018,Buchert_2020}.

In general, $\tilde{\alpha}$ is not a directly observable physical quantity (as it mixes special relativistic angular momentum and curvature) but is rather an expansion parameter, tracking the applicability of post-Newtonian EFT\footnote{The expression for the interaction kernel $R_{\alpha' \beta' \gamma \delta}(x^\mu,y^\mu)$ reduces to $\left[R_{\alpha \beta \gamma \delta}(t; \vec{x}) + R_{\alpha \beta \gamma \delta}(t; \vec{y})\right]/2$ at first order within the post-Newtonian EFT, simplifying the calculation of the expansion parameter.} by sampling relativistic angular momentum over scales comparable with 
curvature variations. 

However, it can be roughly evaluated for astrophysical systems via observables that allow for estimates of their global curvature, $\ave{R}$, (e.g., detailed mass modelling and gravitational lensing), and high-resolution observations capable of estimating the volume, $\ave{V}$ (related to the radius of the luminosity distribution), and the special relativistic angular momentum density, $\ave{J}$ (e.g., via velocities reconstructions). Using a rough order of magnitude estimate, we can then obtain an approximated scaling observable for physical systems, i.e.,
\begin{equation}
\label{scaling}
\frac{\tilde{\alpha}}{G} \sim \ave{R}\times \ave{J}^{2} \times \ave{V}^{2} \, .
\end{equation}

The importance of the proposed parameter can then be understood if one considers two fluid volume elements $\dd \Sigma$ and $\dd \Sigma'$ interacting, in special relativity the mutual angular momentum is conserved, so that for any closed loop we have
\begin{equation}
  \label{eq:angmomcons}
 \oint_{\partial \Sigma} J^{\alpha \beta \gamma} \dd \Sigma_\gamma=0 \, .
\end{equation}
This angular momentum conservation, in extended bodies,  leads to vorticity conservation in both extremes of ideal fluid dynamics \citep{Jackiw_2004,Dubovsky_2006} and collisionless kinetic theory \citep{Lynden-Bell_1967}.   The special relativistic virial theorem \citep{Schutz_1985} expressed in the foliation formalism 
\begin{equation}
\label{eq:virial}
\underbrace{\frac{\partial^{2}}{\partial \Sigma_0^{2}} \int \Sigma^{i} \Sigma^{j} T^{0 0}{\text{d}\Sigma_0} }_{\sim V}=2 \underbrace{\int  T^{i j} {\text{d}\Sigma_0}}_{\sim T} \, ,
\end{equation}
is also proven from the conservation, positiveness and boundedness of $T^{\mu \nu}$ as well as the angular momentum conservation as expressed in Eqn.~\eqref{eq:angmomcons}. However, in a curved spacetime, we expect \eqref{eq:angmomcons} to be violated, even in freely falling coordinate systems, by an amount $ \sim \dd \tilde{\alpha}$. In particular, considering local conservation and boundedness at infinity of $T^{\mu \nu}$, it is natural from dimensional analysis to assume that for small values of $\tilde{\alpha}$ we would have 
\begin{align}
& \partial_{\mu} J^{\mu \nu \rho} \sim \order{\tilde{\alpha}^{1/2}} J \, ,\\ &\ave{T}-2\ave{V} \sim \order{\tilde{\alpha}^{1/2}} \times \order{\ave{T},\ave{V}} \, ,
\end{align}
whilst for $\tilde{\alpha} \gg 1$ the post-Newtonian expansion would completely break down. Here, the problem is not the gravitomagnetic effect per se (which was already addressed in detail in \citep{Rowland_2015,Ciotti_2022,Lasenby_2023}) but rather the breaking down of the conservation law of the special relativistic angular momentum together with the nonlocality of its definition. Indeed, for an extended body this means that the special relativistic angular momentum of every volume element w.r.t.\ every other volume element is not conserved. This can then lead to large deviations, even if each pair of volume elements effect is tiny and well described within the post-Newtonian expansion. Characterising the deviation from the conservation law by the pair of volumes is favoured both by the necessity of an adimensional parameter \citep{buckingham} and in analogy with other common physical prescriptions, like the derivation of the Van Der Waals Equation of State \citep{Goodstein_1985} (which also follows the correlation of two volume elements).

A more precise  analogy might be again found in the EFT approach to QCD. Within a medium this is written in terms of Wilson lines, thus introducing important non-locality effects even in ``weak-field'' limits \citep{Wilson_1974}. We note that the $\tilde{\alpha}$ parameter has some similarity to the interaction between two Wilson loops $\ave{W(x) W(x')}$ in QCD.  
\begin{equation}
\ln W= \mathrm{Tr}_a \oint A^{a}_\mu dx^\mu \propto \mathrm{Tr}^a \left[ F^{\mu \nu}F_{\mu \nu}\right]_a \, ,
\end{equation}
in pure gauge theory this quantity has two scalings w.r.t.\ the coupling constant: weakly coupled perimeter law (which becomes a Coloumb potential) and strongly coupled area law (which denotes confinement) \citep{Greensite_2011}. Close to each limit one can develop an expansion around, respectively, weak and strong coupling, but there is no smooth transition from one to the other.
Continuing our QCD-driven intuition, the Wilson proof of confinement \citep{Wilson_1974} relies on the ``saturation'' of Wilson loops in an area, which parallels the onset of influence of terms proportional to $\tilde{\alpha}$. This suggests that a large $\tilde{\alpha}$ means that a weak-field expansion is inadequate, and a post-Newtonian theory based on an $\tilde{\alpha}$-expansion, analogous to the effective field theory based on Wilson loops and their correlators \citep{Pisarski_2006,Majumder_2013}, becomes necessary.  While applications of this approach to gravity do exist \citep{White_2011,Bonocore_2022}, they are relevant in the vacuum high energy limit, and a systematic EFT expansion relevant to IR many-body physics is still lacking.  Perhaps joining the techniques of \citep{White_2011,Bonocore_2022} with the Wilson line effective theory which was developed in the context of extended bodies in QCD \citep{Pisarski_2006,Majumder_2013} would yield an effective expansion in terms of $\tilde{\alpha}$.

\textcolor{black}{In summary, the key utility of $\tilde{\alpha}$ is that it can be computed within the post-Newtonian limit while providing a quantitative criterion for the breakdown of that limit. Once an effective expansion based on space-like Wilson loops is developed, we expect it to yield an effective theory for extended rotating objects as a series in powers of $\tilde{\alpha}^{-n}$. In the absence of such an expansion, all that can presently be asserted is that the post-Newtonian approximation is reliable when $\tilde{\alpha} \ll 1$, but is expected to fail otherwise. In the next section, we will evaluate $\tilde{\alpha}$ for a range of astrophysical systems where dark matter has been claimed to play a significant role.
}
\section{Estimation for astrophysical systems}
To test the validity of our reasoning, we set out to calculate the value of $\tilde{\alpha}$ for different astrophysical systems of interest, within the conventionally accepted Newtonian approximation of GR\footnote{We note that employing the Newtonian approximation to estimate $\tilde{\alpha}$ does not constitute a contradiction w.r.t. the scope of the parameter itself as the purely GR components of $\tilde{\alpha}$ can well be approximated by the usual post-Newtonian expansion in the right limit. It is the dynamic of the system which might not then be well approximated by the post-Newtonian expansion when $\tilde{\alpha}$ calculated this way is large.}. We present an estimate for the value of $\tilde{\alpha}$ for the Alpha-Centauri A and B stellar binary \citep{Wiegert_1997}, a pulsar-neutron star binary system (the Hulse–Taylor pulsar, also identified as PSR B1913+16 \citep{Taylor_1982})\footnote{We note that, in QCD, we can find the equivalent for a binary system, i.e., a quarkonium state \citep{Brambilla_2000}, where the effective theory is written in terms of a two-body potential.}, a stellar globular cluster, a dwarf disc galaxy, a Milky Way-like disc galaxy, \textcolor{black}{a dark matter-deficient ultra-diffuse elliptical galaxy (FCC226, see e.g., \cite{Munoz_2015,Venhola_2017})}, a massive Elliptical galaxy, and \textit{our home in the Cosmos}, the super-cluster Laniakea \citep{Tully_2014,Valade_2024}. We note that if the scalar parameter $\tilde{\alpha}$ correctly traces the onset of nonlocal angular momentum effects within GR, i.e., the possible breakdown of the post-Newtonian approximation, its value must be much smaller than unity for all binary systems, where the post-Newtonian approximation has been directly tested \citep{Poisson_2014,Clifford_2014}.

To model the binary systems \citep{Wiegert_1997,Taylor_1982}, we take the limit of point-like bodies\footnote{To avoid divergence in the Newtonian potential we employ within the astronomical objects the effective gravitational potential $\Phi(r) = -GM\left(3r_b^2-r^2\right)/2r_{b}^3$, where $r_b$ represents the radius of the objects, much smaller than the typical orbital radius. Such a choice can be understood in a rough modelling of the astronomical bodies as spheres of uniform densities.} in stationary circular orbits w.r.t.\ their centre of mass. Moreover, we employ the standard, static Plummer mass profile for the globular cluster \citep{Plummer_1911,Binney_2008}, whilst to model the baryonic component of the disc galaxies we employ a conventional axisymmetric, stationary, razor-thin disc mass distribution, i.e., the Kuzmin profile \citep{Kuzmin_1956,Binney_2008}. Finally, to model Laniakea, we consider its recent determination \citep{Dupuy_2023}\footnote{In this paper, we do not employ the most recent determination of Laniakea \citep{Valade_2024} as no velocity field coupled to a density field reconstruction is available at the moment.}, which employs the velocity and density field reconstruction of \citep{Courtois_2023}, derived from
the latest Cosmicflows-4 (CF4) data release \citep{Tully_2023}. Thus, in our paper Laniakea is represented by 4079 voxels within a grid of 128$^3$ cells covering an overall volume of 1 (Gpc/$h)^3$, where $h$ is the reduced Hubble constant.

To estimate $\tilde{\alpha}$, we consider the conventional Newtonian line element of GR in cylindrical coordinates for all the systems but the elliptical galaxy and Laniakea, for which it is convenient to use standard Cartesian coordinates. Thus, \textcolor{black}{for all the systems}, we \textcolor{black}{employ the following metric}
\begin{align}
    \label{eq:Nmetric}
    \dd s^2 = &-\left(1-\frac{2\Phi}{c^2}\right)c^2\dd t^2 + \left(1+\frac{2\Phi}{c^2}\right)\textcolor{black}{\gamma_{ij}\dd x^i\dd x^j}\, , 
\end{align}
where $\Phi(t,\vec{x})$ is the Newtonian potential, \textcolor{black}{and $\gamma_{ij} = \mathrm{diag}(1,1,1)$ in cartesian coordinates and $\gamma_{ij} = \mathrm{diag}(1,1,r^2)$ in cylindrical}. The geometry of Eqn.~\eqref{eq:Nmetric} is then coupled to the energy-momentum tensor $T^{\mu\nu}$ of these systems, \textcolor{black}{i.e,} 
\begin{equation}
    \label{eq:EMT} T^{\mu\nu}=\left(\rho+p/c^2\right) \, U^{\mu}U^{\nu}\, + p g^{\mu\nu}\, ,
\end{equation}
where $\rho(t,\vec{x})$ is the matter density, $p(t,\vec{x})$ is the matter internal pressure and $U^\mu$ is the four-velocity of the  body/fluid element. \textcolor{black}{For the axisymmetric systems this is} given by
\begin{equation}
    \label{eq:4vel}
    U^\mu\partial_\mu = \frac{1}{\sqrt{-H}}\left(\partial_t + \Omega \partial_\phi\right)\, ,
\end{equation}
where $\Omega(t,r,z,\phi)$ is the point-wise defined rotational velocity of the fluid elements/orbiting bodies, and $H(t,r,z)$ is the normalisation factor fixed by the condition $U^\mu U_\mu = -c^2$, i.e.,
\begin{equation}
    H(t,r,z,\phi) = -1 + \frac{2\Phi}{c^2} + \left(\frac{r\Omega}{c}\right)^2\left(1 + \frac{2\Phi}{c^2}\right)\, ,
\end{equation}
\textcolor{black}{On the other hand, for both the elliptical galaxy and Laniakea, the four velocity is a function of all three spatial coordinates.} We can now specialise Eqs.~\eqref{eq:EMT} and \eqref{eq:4vel} to the different systems under study. The binary systems, being only rotationally supported, are taken with a zero internal pressure, a constant orbital rotational velocity, i.e. $\Omega(r,z) = \Omega_0$, and a density profile given by
\begin{equation}
\label{eq:dens_bin}
    \rho(t,r,z,\phi) = \frac{\bar{M}}{r}\delta(z)\delta(r-r_{0})\left[\delta(\phi - \Omega_0 t) + \delta(\phi - \Omega_0 t + \pi)\right] \, ,
\end{equation}
where $\bar{M}$ is the average mass of the two bodies in the system, $\delta$ is the Dirac delta distribution, $\Omega_0$ is their orbital rotational speed and $r_{0}$ is the radius of the orbit. We take $\bar{M} = 1\,\text{M}_\odot$ and $\bar{M} = 1.4\,\text{M}_\odot$ for the stellar binary and the Hulse–Taylor pulsar, respectively. Moreover, we have $\Omega_0 = 2.5 \cdot 10^{-9}\, \text{rad/s}$ and $\Omega_0 = 2.25 \cdot 10^{-4}\, \text{rad/s}$, respectively, whilst $r_{0} = 24\, \text{AU}$ and $r_{0} = 0.01\, \text{AU}$, respectively, for the two systems. For the stellar globular cluster, we are considering a static effective fluid description derived from a time-average over the chaotic orbits of the stars within the globular cluster, such that $\Omega(r,z) = 0$, and the density is given by spherically symmetric distribution \citep{Plummer_1911,Binney_2008}
\begin{equation}
    \rho(r,z) = \frac{3\,\Rb^2\Mb}{4\pi\left(r^2+z^2+\Rb^2\right)^{5/2}} \,,
\end{equation}
where $\Rb$ is the core radius of the globular cluster, so that the whole cluster should be contained within two to three such radii, and $\Mb$ is its total mass. In our estimates, we model a massive globular cluster, and we thus take the realistic values $\Rb = 30 $ pc and $\Mb = 10^6\,\text{M}_\odot$. For both disc galaxies, the density profile is given by the Kuzmin profile \citep{Kuzmin_1956,Binney_2008}, i.e., 
\begin{equation}
\label{eq:rdensity}
    \rho(r,z)=  \frac{\Md \rd}{2\pi\left(r^2 +
    \rd^2\right)^{3/2}}\delta(z)\, ,
\end{equation}
where $\Md$ is the total disc mass and $\rd$ is the scale length in the radial direction for the galactic disc. We take $\Md = 10^{6} \,\mathrm{M}_\odot$, $\Md = 10^{10} \,\mathrm{M}_\odot$, and  $\rd = 0.5 \,\mathrm{kpc}$, $\rd = 2.8 \,\mathrm{kpc}$, respectively, for the dwarf disc galaxy and the Milky Way-like disc galaxy.  The angular velocity field of the fluid elements for the disc galaxy model is then given by
\begin{equation}
    \Omega(r,z) = \delta(z)\sqrt{\frac{G\Md}{\left(r^2+\rd^2\right)^{3/2}}}\, .
\end{equation}
We can also directly define the Newtonian potentials for the two extended matter configurations with an analytical description for their densities (we note that for the binary systems, this is just the superposition of the potential generated by two point-like sources). For the globular cluster we have
\begin{equation}
    \Phi(r,z) =  -\frac{G\Mb}{\sqrt{r^2+z^2+\Rb^2}}\,,
\end{equation}
and for the disc galaxies
\begin{equation}
    \Phi(r,z) = -\frac{G\Md}{\sqrt{r^2+(|z|+\rd)^2}}\,.
\end{equation}
Furthermore, the pressure can be obtained numerically by solving for hydrostatic equilibrium in these systems \citep{Binney_2008}. \textcolor{black}{Here, we note that by solving for hydrostatic equilibrium we do not need to select an equation of state as the pressure is directly solved for, without any approximation.} For the elliptical galaxies we instead consider a static, triaxial Dehnen model \citep{Binney_2008} with 
\begin{equation}
    \rho(x,y,z) = \frac{3\gamma M}{4\pi abc}\left(m^\gamma(1+m)^{4-\gamma}\right)^{-1} \, ,
\end{equation}
where $a,\, b$, and $c$ are the scale-lengths of the principal axis, $M$ is the total mass, $\gamma$ is the inner slope parameter, and $m:= ((x/a)^2 + (y/b)^2 + (z/c)^2)^{1/2}$.  The respective Newtonian potential is given by \citep{Binney_2008}
\begin{equation}
    \Phi(x,y,z) = - G\int_0^\infty \frac{M(m_\lambda)}{\sqrt{(a^2+\lambda)(b^2+\lambda)(c^2+\lambda)}}\dd \lambda \,
\end{equation}
where $m_\lambda:=(x^2/(a^2+\lambda) + y^2/(b^2+\lambda) + z^2/(c^2+\lambda))^{1/2}$, and $M(m_\lambda) := Mm_\lambda^{3-\gamma}/(1+m_\lambda)^{3-\gamma}$. The velocity dispersion, i.e., the marker of effective pressure in elliptical galaxies, can then be found as for the other models, \textcolor{black}{i.e., by solving for hydrostatic equilibrium}. Here, to estimate $\tilde{\alpha}$, \textcolor{black}{we take for the ultra-diffuse galaxy $M = 2\times10^{8}$ M$_\odot$, $a = 8.1$ kpc, $b = 6.9$ kpc, $c = 5.7$ kpc, and $\gamma = 0.2$, whilst for the massive elliptical} $M = 5\times10^{11}$ M$_\odot$, $a = 4.5$ kpc, $b = 2.7$ kpc, $c = 2.3$ kpc, and $\gamma = 1.5$. We then discretised these models on a three-dimensional grid of $128^3$ cells, covering the volume for the elliptical galaxies. We can thus compute numerically the Riemann tensor on the grid, as well as the other physically relevant quantities.

On the other hand, the energy-momentum tensor for Laniakea is directly obtained from the three-dimensional density and velocity field reconstructions \citep{Courtois_2023,Dupuy_2023} from CF4 \citep{Tully_2023}, whilst the Newtonian potential is numerically derived from the superposition of the Newtonian potential generated by each voxel in the reconstruction. \textcolor{black}{Here, as standard practice in cosmology, we model Laniakea as being pressureless, i.e., dust. The presence of pressure support is taken into account only at the level of the non-trivial velocity dispersion within the system.} Moreover, to estimate the $\tilde{\alpha}$ parameter, we consider Laniakea as an isolated system w.r.t.\ to the rest of the Universe, so we fix the density field outside Laniakea to be zero. Therefore, following the same procedure as in \citep{Giani_2024}, we subtract from the reconstructed velocity field the Hubble flow to offset the effects of any external gravitational field.

We can now calculate the parameter $\tilde{\alpha}$ for these systems at leading order in powers of $\epsilon := v/c$, where $v$ is the typical nonrelativistic velocity in these systems. We note that under the Newtonian approximation, we also have $\Phi/c^2 \goesas \epsilon^2$. \textcolor{black}{Hence, the Riemann tensor is described at leading order as \citep{Poisson_2014}
\begin{equation}
    R_{\alpha \beta \gamma \delta} = \delta_{\alpha 0}\delta_{\gamma 0}\partial_\beta\partial_\delta\Phi + \mathcal{O}(\epsilon^3)\, ,
\end{equation}
i.e., the Riemann tensor reduces to the tidal tensor of the underlying Newtonian potential. We therefore see that if we define the infinitesimal angular momentum tensor 
\begin{equation}
    \dd\mathcal{M}^{\mu\nu} : = J^{\mu\nu\rho}\dd\Sigma_\rho \, ,
\end{equation}
the calculation for the $\tilde{\alpha}$ parameter will single out the time-space components of the relativistic angular momentum, $\dd\mathcal{M}^{0i}$. Therefore we have
\begin{equation}
    \tilde{\alpha} = G \int\int_{V \times V} R_{0'i'0j}\dd\mathcal{M}^{0'i'}\dd\mathcal{M}^{0j} \, .
\end{equation}
After some rather lengthy algebra we find for all axisymmetric systems}\footnote{We have submitted our Mathematica code, \lq\lq \textit{Nonlocal\_Expansion\_GR}'' as an ancillary file. The code defines the spacetime metric at first order in the post-Newtonian formalism, and calculates the relative energy-momentum tensor in cylindrical coordinates for the matter source. Finally, it obtains at first order in the expansion the formula for $\tilde{\alpha}$, i.e., Eq.\ \eqref{eq:alpha_final}.}
\begin{align}
     \label{eq:alpha_final}
     \tilde{\alpha} \simeq& -\frac{16\pi^2 G}{c^2}\iiiint r\rho(r,z)r'\rho(r',z')\left[rr'\partial^2_r\Phi(r,z) + zz'\partial^2_z\Phi(r,z)\right.    \nonumber\\  & +   \left. \left(rz' + zr'\right)\partial_r\partial_z\Phi(r,z)\right] \dd r \dd r'\dd z \dd z' \, ,
\end{align}
where for the binary systems the $\phi$ dependence has already been integrated out, thus defining an effective $\rho(r,z)$ for such systems. For the elliptical galaxies and Laniakea, since we employ data on a three-dimensional grid, Eq. \eqref{eq:myalpha} is discretised and evaluated at the leading order. Moreover, we point out that given the reasoning used in the definition of $\tilde{\alpha}$ in terms of vorticity conservation and virial theorem,  $G \times M_\odot^2$ is the natural scale to measure $\tilde{\alpha}$ since it matches the scale of the ``atoms'' (i.e. stars) making up the astrophysical systems under considerations. The results are summarised in Table \ref{tab:1}\textcolor{black}{, alongside the volume\footnote{\textcolor{black}{We omit the volume for binary systems, here modeled as essentially orbiting point-particles.}} and the mass of the systems under considerations}.
\begin{table*}[ht]
\renewcommand{\arraystretch}{1.5}
\centering
\caption{Estimates for the values of $\tilde{\alpha}$ for the astrophysical systems considered.}
\begin{tabular}{|c|c|c|c|}
\hline
 & \textcolor{black}{Volume [kpc$^3$]} & \textcolor{black}{Mass [M$_\odot$]} & Estimate for the $\tilde{\alpha}$ parameter \\
\hline
Stellar binary & \textcolor{black}{/} &   \textcolor{black}{1} & $10^{-9}$ \\
\hline
Pulsar binary  & \textcolor{black}{/} &  \textcolor{black}{$1.4$}  & $10^{-5}$ \\
\hline
Globular Cluster & \textcolor{black}{$\approx10^{-4}$} & \textcolor{black}{$10^{6}$}  & $10^{-4}$ \\
\hline
Dwarf disc galaxy & \textcolor{black}{$\approx10^{-2}$} &    \textcolor{black}{$10^{6}$} & $10^{4}$ \\
\hline
Massive disc galaxy & \textcolor{black}{$\approx5$} &  \textcolor{black}{$10^{10}$}   & $10^{13}$ \\
\hline
Ultra-diffuse galaxy & \textcolor{black}{$\approx10^3$} &  \textcolor{black}{$2\times10^{8}$}   & $10^{-1}$\\
\hline
Elliptical galaxy & \textcolor{black}{$\approx10^{2}$} &  \textcolor{black}{$5\times10^{11}$}   & $10^{9}$\\
\hline
Laniakea & \textcolor{black}{$\approx10^{18}$} &     \textcolor{black}{$\approx10^{17}$} & $10^{26}$ \\
\hline
\end{tabular}
\label{tab:1}
\end{table*}

We find that the results in Table \ref{tab:1} are in line with the expectations following our reasoning. The novel parameter we propose takes fairly different values whether we consider isolated binaries or continuous, spinning bodies. The values for $\tilde{\alpha}$ are well below unity in systems for which the post-Newtonian has been directly verified. Moreover, the same is true for the globular cluster, in which the static assumption -- which relates to the isotropic average zero velocity of the effective fluid element in such systems -- implies that only the exclusively special relativistic components of the angular momentum contribute to the estimate of $\tilde{\alpha}$, giving overall a small contribution. \textcolor{black}{Furthermore, for the ultra-diffuse elliptical galaxy, we find $\tilde{\alpha} \approx 0.1$, suggesting only minor corrections to a post-Newtonian dynamical description of the system. This result again illustrates how low relativistic angular momentum—corresponding to small energy gradients and velocities—can yield $\tilde{\alpha}$ values well below unity, even for spatially extended bodies.}

In contrast, for wide-extended bodies with a considerable relativistic angular momentum over a scale where the curvature varies, such as a disc galaxy or Laniakea, the obtained value is orders of magnitude above unity. In this regard, the $\tilde{\alpha}$ parameter spans eight orders of magnitude between the massive globular cluster and the dwarf disc galaxy, having the same total mass. Such a difference cannot be reconciled with simply a different spatial extension of the two systems, \textcolor{black}{as the volume of the two system differs only by two orders of magnitude.} Instead, it signals a direct difference in the overall degrees of freedom in the relativistic angular momentum distributions of the two systems.

Moreover, although the massive elliptical galaxies and superclusters are predominantly pressure-supported systems, we still obtain a value of $\tilde{\alpha} \gg 1$. This can be understood by recognising that the relativistic angular momentum entering the definition of our proposed expansion parameter accounts not only for rotational motion, but also for pressure gradients and spatial variations in the energy density (see e.g., \citep{Gourgoulhon_2012}). \textcolor{black}{This can be understood by noting that the relativistic angular momentum is defined in terms of the full energy-momentum tensor of the system, where the pressure contributes both directly, as a source, and indirectly, by shaping the velocity dispersion and density gradients (and hence gravitational gradients)
through hydrostatic equilibrium.}

As argued in this paper, a value for the $\tilde{\alpha}$ parameter above unity could then signal a break-down of the post-Newtonian approximation -- whose applicability is no longer guaranteed by the KAM theorem \citep{Villani_2011} -- and thus the necessity for a novel EFT appropriate for $\tilde{\alpha} \gg 1$, perhaps by using approaches such as \citep{White_2011,Bonocore_2022}. Furthermore, the values of $\tilde{\alpha}$ below and above unity correlate with the inferred absence, or presence of abundant dark matter in the astrophysical systems. Hence, it is reasonable that a nonlocal EFT of GR might impact dark matter estimates in cosmological and astrophysical settings.

\section{Conclusions}

Our considerations in this paper are very generic, anticipating the construction of a novel EFT of GR. They should apply to any situation where there is non-negligible angular momentum (defined in special relativistic terms) over scales ``large'' compared to the curvature. At the moment, we do not have an effective theory in $\tilde{\alpha}$ which can be used to actually calculate corrections to the post-Newtonian expansion. 

However, we can discuss where such effective theory corrections might become necessary. \textcolor{black}{While galaxy rotation curves are the leading historical evidence for dark matter, it is nowadays far from the only one.  However, as we shall now briefly describe, all other pieces of evidence concerns regimes where each element of \eqref{scaling} is non-trivial, and hence where $\tilde{\alpha}$ is likely to be non-negligible.} 

For instance, let us consider the dynamics of galaxy clusters, in the light of the value of the parameter obtained for $\tilde{\alpha}$ for Laniakea. In particular, the Bullet cluster \citep{Clowe_2004}, also fits in the range of validity of the approach here, since it is a collision with significant velocities and an impact parameter comparable to the radius of curvature of the galaxies. The calculations here do not touch the issue of light-bending in relativity, but it is reasonable that a nonlocal expansion around a scale such as $\tilde{\alpha}$ could yield significant corrections to the lensing observables used to characterise the dark matter distribution within galaxy clusters \citep{Clowe_2004,Bartelmann_2010}. Moreover, we note that in the original work by Deur \citep{Deur_2009, Deur_2017}, nonlocal effects of GR -- discussed in analogy with self-interactions of non-abelian effective field theories -- already showed non-negligible impact on galaxy rotation curves. \textcolor{black}{What this means is that care needs to be taken when discussing the \lq\lq non-relativistic limit''.  Usually, such a limit is defined by small velocities and weak purely time-like metric potentials, but as we see this might not be enough: the Riemann tensor is sensitive to space gradients of such potentials, and if an angular momentum is defined over regions where such gradients are large, its violation can not be captured by the usual post-Newtonian expansion at leading orders (see also \cite{Buchert_2009} for a complementary discussion).  For binaries most of this violation goes away when reduced coordinate systems are chosen, but such choice makes no difference if the number of gravitating bodies is much bigger than two.  Because of this the effective theory of general relativity must include non-local parameters, of which $\tilde{\alpha}$ is expected to be the leading candidate (as, to use the terminology from particle effective theory, it is the lowest dimensional operator)}

Most importantly, the relevance of the suggested nonlocal expansion of GR, and its relation to dark matter estimates, can be directly tested with the upcoming optical surveys, i.e., Euclid, DESI and LSST. \textcolor{black}{Indeed, the expansion parameter $ \tilde{\alpha}$ has both a physical picture associated with it and it is based on quantities measurable independently (see Eq.\ \eqref{scaling}). Hence, the validity of the scaling of $\tilde{\alpha}$ w.r.t. dark matter abundance across different scales (clusters, small and large galaxies, galaxy clusters, regions of the universe,...) gives a quantitative and falsifiable test on whether the presence of hitherto neglected general relativistic corrections for many-body systems with non-negligible angular momentum is part of the resolution of the dark matter problem. Specifically,} the scaling can be verified using detailed galaxy surveys that probe the necessary astrophysical quantities. In this regard, we note that ultra-faint dwarf galaxies \citep{Simon_2019} would constitute an ideal probe as their large morphological variance (within a well-defined low mass range) will allow upcoming surveys to gauge a wide range of the observables of interest. Indeed, galaxies with large $\ave{R}$ but smaller $\ave{J},\ave{V}$ should exhibit a smaller dark matter fraction than galaxies with similar curvature but larger $\ave{J},\ave{V}$. With a sample with a broad distribution in $\ave{R},\ave{J},$ and $\ave{V}$ separately, observations should distinguish between a robust scaling (that would provide evidence that a novel, nonlocal EFT of GR wold be relevant for dark matter estimates) from a large spread (which would prove that dark matter estimates are unrelated to the expansion given here). 

Finally, one piece of evidence of dark matter where angular momentum seems to play no role is based on the anisotropies of the CMB \citep{Planck_2016}.  However, if the evidence of magnetic fields on such scales \citep{Matarrese_2005,Durrer_2013} is vindicated, independently of their origin, an angular momentum associated with them is inevitable, and a non-local general relativistic effect arising from this angular momentum could be non-negligible. Moreover, the work by Bruni and collaborators \citep{Bruni_2013,Thomas_2015} directly points importance that angular momentum and frame-dragging might have even in cosmology at the level of structure formation, \textcolor{black}{whose rapid dynamics is also counted as evidence for dark matter}.

In addition to empirical scaling investigations, our ideas could in principle be assessed by numerical relativity.   So far, hydrodynamic simulations with full relativity have concentrated on few-body problems such as mergers.  However, the technology to simulate axisymmetric solutions with many degrees of freedom (such as fluids) is in principle available \cite{rezzolla}.   It should therefore be possible to check if there is a bigger-than expected deviation from the post-Newtonian paradigm related to our scaling observable.

In conclusion, driven by analogies with non-Abelian gauge theory, we have argued that post-Newtonian effective field theory might fail when the proposed adimensional constant $\tilde{\alpha}$ is non-negligible. We have estimated $\tilde{\alpha}$ for a variety of astronomical objects, and found that it is small for systems well described by the post-Newtonian approximation and with no signs of dark matter, while it is large for disc galaxies and Laniakea, where dark matter is thought to play an important role. We point out that potentially $\tilde{\alpha}$ could be non-negligible for other setups, such as the Bullet cluster and cosmological scales, where dark matter dominates the dynamics. Therefore, the future development of a nonlinear many-body effective theory of general relativity suitable for large $\tilde{\alpha}$ represents an interesting challenge. Its precise role in dark matter estimates, and thus its relevance in modern cosmology, could be ascertained with a survey of galaxy properties, particularly ultra-faint dwarf galaxies, by verifying if the perceived dark matter component scales with the estimate of $\tilde{\alpha}$ from observations.
\\
\\
\section*{Acknowledgements}
M.G. acknowledges partial support by the Marsden Fund administered by the Royal Society of New Zealand, Te Apārangi under grant M1271. G.T. acknowledges support from Bolsa de produtividade CNPQ 305731/2023-8, Bolsa de pesquisa FAPESP 2023/06278-2. The authors are deeply grateful to Robert Brandenberger, Paul Chouha, Pierre Mourier, Syksy Räsänen, and Federico Re for their insight and our fruitful discussions via correspondence. We are deeply thankful to Riccardo Sturani for his clear comments and the opportunity to present this work at the ICTP-SAIFR in São Paulo. We sincerely thank Zachary Lane for his irreplaceable expertise in dealing with observational data. The authors are thankful to Leonardo Giani for his insights on the analysis carried out for our home in the cosmos, Laniakea. We are grateful to Luca Ciotti for the meaningful discussions on the testable predictions of our framework. We thank Christopher Harvey-Hawes, Morag Hills, Emma Johnson, Shreyas Tiruvaskar, Michael Williams and David Wiltshire for useful discussions.

\bibliographystyle{elsarticle-harv}
\bibliography{references.bib}

\end{document}